\documentclass[conference]{IEEEtran}

\ifCLASSINFOpdf
  \else

\fi

\usepackage{amsmath}
\usepackage{graphicx}
\usepackage{caption}
\usepackage{tabularx,multirow,booktabs,lipsum}
\usepackage{subfig}
\usepackage{color}
\usepackage{float} 
\usepackage{paralist} 
\usepackage[T1]{fontenc}
\usepackage{currvita}
\usepackage{epstopdf}
\usepackage{amsmath}
\usepackage{amsfonts}
\usepackage{amssymb}

\begin{document}

\title{Secure and Robust Authentication for DC MicroGrids based on Power Talk Communication}

\author{\IEEEauthorblockN{Marko Angjelichinoski, Pietro Danzi, \v{C}edomir Stefanovi\'c, Petar Popovski}
\IEEEauthorblockA{Department of Electronic Systems, Aalborg University, Denmark \\
Email: \{maa,pid,cs,petarp\}@es.aau.dk }
}

\maketitle

\begin{abstract}
We propose a novel framework for secure and reliable authentication of Distributed Energy Resources to the centralized secondary/tertiary control system of a DC MicroGrid (MG), networked using the IEEE 802.11 wireless interface.
The key idea is to perform the authentication using \emph{power talk} -- a powerline communication technique executed by the \emph{primary control loops} of the power electronic converters. 
In addition, the scheme also promotes direct and active participation of the control system in the authentication process, a feature not commonly encountered in current networked control systems for MicroGrids.
The PLECS\textsuperscript{\textregistered{}}-based simulations verifies the viability of the proposed solution.
\end{abstract}

\IEEEpeerreviewmaketitle

\section{Introduction}

In recent years, the traditional AC power grid has been subjected to unprecedented proliferation of flexible, small-scale Distributed Energy Resources (DERs) and smart loads, many of which are direct current (DC) in nature, leading to formation of self-sustainable low voltage DC clusters referred to as \emph{MicroGrids (MG)}.
In contrast to AC distribution systems, DC MGs provide significant installation, operation and maintenance flexibility due to high penetration of renewable DERs that are interfaced to the DC infrastructure through \emph{power electronic converters} {\cite{ref1m,ref2m}}.
Specifically, the extensive use of power electronic converters provides for implementation of advanced control and optimization mechanisms in DC MGs, where the feedback loop is typically closed via \emph{external communication network} \cite{ref01}. 

The ongoing trend emphasizes the importance of the security of the control system, particularly the security of the feedback loop, and raises concerns regarding the adequacy of existing communication technologies for the security challenges of the evolving DC MG ecosystem {\cite{ref00}}.
Specifically, the control feedback loop is commonly closed via off-the-shelf communication solutions such as IEEE 802.11 \cite{ref02}.
Even though the current {security specification} of the standard, namely the IEEE 802.11i, is considered to be safe, a potentially vulnerable point of failure in the security system is the \emph{initial handshake procedure}, which might prevent the {DER units that aim to join the control system} from accessing the communication resources, resulting in poor regulation and even instability.
In all current MG implementations, authentication by the communication system automatically grants access to control system information, i.e., access to control.
This situation might lead to undesired and potentially harmful circumstances in existing and emerging MG systems, especially in tactical or military applications, where a malicious entity may obtain access to the communication system and thus compromise MG operation.
So far, there has been very little effort to design secure control architecture in which the control system actively participates in authentication and access control.

This paper proposes novel, robust networked control architecture for secure authentication of DER units, that employs recent powerline communication (PLC) technique termed \emph{power talk}, and runs the initial handshake over the powerlines.
Power talk {\cite{ref3m,ref4m,ref5m}} is an in-band, low rate PLC solution, designed originally for self-sustainable DC MGs that do not rely on external communication systems.
It modulates the information into subtle deviations of the parameters of the \emph{primary droop control} of the DERs, 
which translate in information-carrying deviations of the steady state voltage of the DC distribution system.
Using power talk, the handshake becomes effectively invisible for the conventional attacker, as the attacker needs to physically access the grid in order to perform the attack, which is significantly more complicated and often impossible (e.g., MG systems for applications guarded by a safety perimeter).
Furthermore, the proposed scheme promotes active participation of the MG control system in the authentication procedure, as it is embedded within the primary control loops, without the use of any additional, external hardware, requiring only software modifications of the power electronic converters.

The paper is organized as follows.
Section~\ref{sec:security} reviews the state-of-the art security procedures used in IEEE 802.11 and discusses their potentially weak points, with the emphasis on the handshake procedure.
Section~\ref{sec:control} introduces the hierarchical MG control architecture, discusses possible cyber-attacks and introduces the basics of power talk.
Section~\ref{sec:main} presents the secure authentication procedure.
Section~\ref{sec:results} verifies the viability of the advocated solution through PLECS\textsuperscript{\textregistered{}} simulation of a realistic low voltage DC MG. 
Section~\ref{sec:conc} concludes the paper.

\section{Cryptographic Handshake of IEEE 802.11 systems}
\label{sec:security}

\begin{figure}[!tb]
\centering
{\includegraphics[width=0.5\columnwidth]{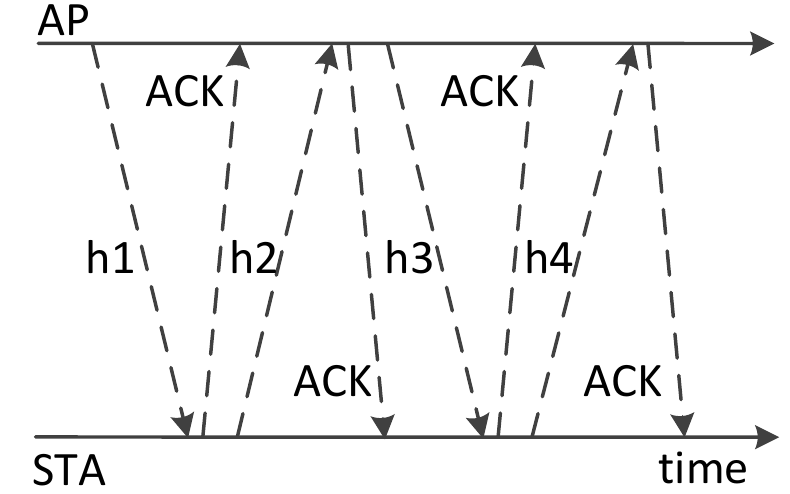}}
    \caption{Representation of the four-way-handshake.}
    \label{fig:hs}
\end{figure}

The IEEE 802.11 standard includes advanced security mechanisms to ensure the confidentiality and integrity of the communication channel.
The most recent security specification is IEEE 802.11i \cite{ref06}, whose security aspects are summarized as follows.

The standard ensures the confidentiality and integrity of the packets by means of the CCMP protocol, that is based on the Advanced Encryption Standard (AES) cipher.
A shared key, named Pairwise Transient Key (PTK), guarantees the encryption between Station (STA) and Access Point (AP).
A second shared key, the Group Transient Key (GTK), is used for multicast and broadcast traffic.
The STA is provided with the keys after having proved to be authorized, i.e., to know the right Pairwise Master Key (PMK).
The PMK can be derived from a Pre-Shared Key (PSK) or by means of advanced authentication mechanisms that involve an Authentication Server.
In any case, STAs have to indirectly prove the knowledge of the PMK to the AP; this problem has been resolved by means of a \emph{cryptographic handshake} \cite{ref04}.
The handshake consists of four messages, each requiring an Acknowledgement (ACK).
The protocol steps, depicted in Fig.~\ref{fig:hs}, can be summarized as: (\emph{$h_1$}) AP$\rightarrow$STA containing the ANonce and its own MAC address enabling the STA to compute the PTK, (\emph{$h_2$}) STA$\rightarrow$AP containing the CNonce and its own MAC address, enabling the AP to compute the PTK, (\emph{$h_3$}) AP$\rightarrow$STA containing the GTK, and, (\emph{$h_4$}) STA$\rightarrow$AP confirmation of the reception.

Besides the basic dictionary attack \cite{ref05}, particularly threatening attack on the handshake is when a malicious attacker impersonates the STA and sends invalid packets leading to unsuccessful authentication \cite{ref04} and disabling the STA from accessing the communication resources.
This situation may result in poor regulation of the MG system, leading to performance degradation and instability as the DER is prevented from communicating and, thus, participating in the control and optimization, c.f. \cite{ref03}.

\section{Control in DC MicroGrids}
\label{sec:control}

\subsection{Multiple-Bus DC MicroGrid Architecture}

A DC MG is a collection of DERs and loads, connected to distribution infrastructure that comprises set of buses interconnected via distribution lines.
The total number of DERs is denoted with $U$, and they are indexed in the set $\mathcal{U}=\left\{0,...,U-1\right\}$.
They use power electronic converters to interface the distribution system, and their voltage and current (i.e., power) outputs are locally regulated via several control channels of different bandwidths.
{The total demand of the aggregate load is denoted with $d$.}

\subsection{Hierarchical Control}
\label{sec:hc}

\begin{figure*}[!t]
\centering
{\includegraphics[width=0.99\textwidth]{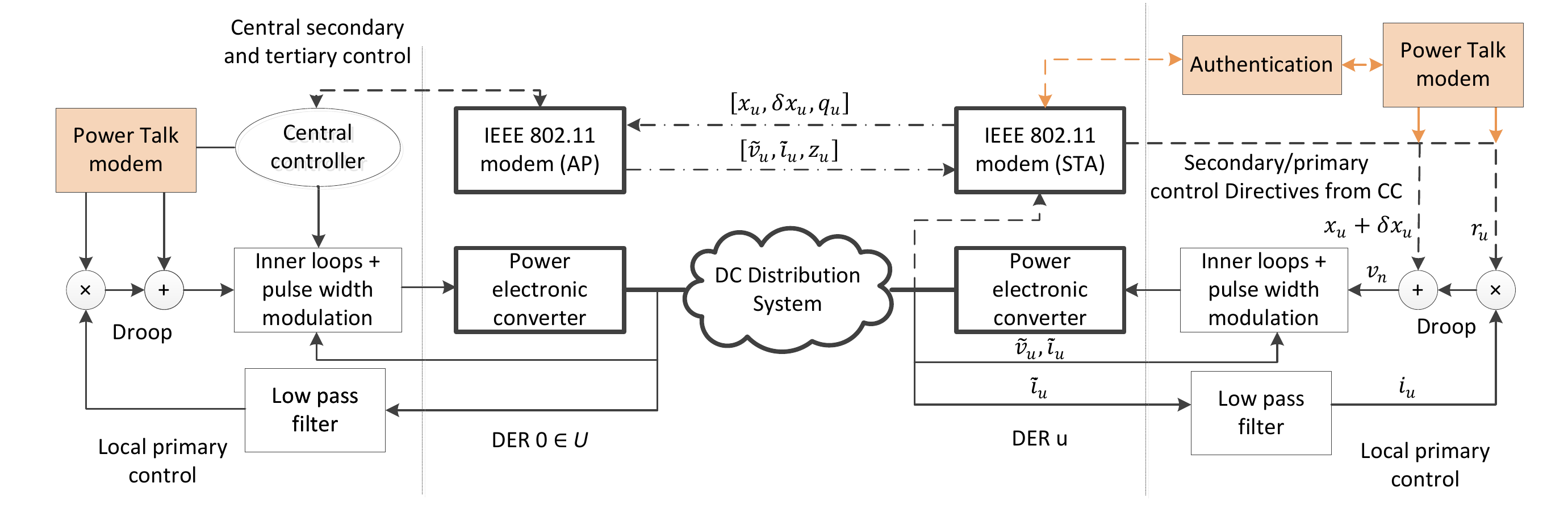}}
    \caption{Control diagram and flow of information in DC MG with centralized secondary/tertiary control and secure and reliable power talk based DER authentication. The red blocks are the newly added software components.}
		\label{fig:control}
\end{figure*}

The control architecture of the MG is summarized in Fig.~\ref{fig:control}.
The control plane is organized in a hierarchy, comprising fast decentralized primary control, and slow centralized secondary and tertiary control.\footnote{We adopt the centralized secondary/tertiary architecture for simplicity of exposition. However, we note that the proposed solution can be easily applied to the distributed case \cite{ref02} with minor modifications.}
In standard implementations, the primary control uses only local measurements as feedback, i.e., it does not require any exchange of information among remote units.
On the other hand, the feedback loop of the secondary/tertiary control is closed via external communication system; thereby, each power electronic converter is assumed to be pre-equipped with IEEE 802.11 modem, see Fig.~\ref{fig:control}.
We assume: (i) the communication is centralized, i.e., there is a single AP and all DERs are associated with it, 
and (ii) the central secondary/tertiary controller (CC) is collocated with the AP in the same physical unit, see Fig.~\ref{fig:control}.
Thus, besides handling the secondary/tertiary control processes, the same unit is also in charge of the authentication of DERs to the communication network.
Without loss of generality, assume that the secondary/tertiary CC and AP reside in DER $0$.

\subsubsection{Primary control}
A common primary control configuration is in the form of Voltage Source Converter (VSC).
VSC DERs regulate the electrical parameters and balance the supply-demand to guarantee stability, based on local output measurements, using the following law \cite{ref1m,ref2m,ref6m}:
\begin{equation}\nonumber
	v_u = x_{u} - r_u i_{u},\;u\in\mathcal{U},
\end{equation}
where $v_u$ and $i_{u}$ are the bus output voltage and current, $x_{u}$ is the \emph{reference voltage} and $r_u$ is the \emph{virtual resistance}.
This implementation is known as \emph{decentralized droop control}.
The value $v_n$ serves as input reference to the inner primary control loops, that operate with frequency $\nu$, equal to the sampling frequency of the converters' ADC, see Fig.~\ref{fig:control}.
The droop controller controls $x_{u}$ and $r_u$, where $x_{u}$ determines the voltage rating of the system, while the $r_{u}$ determines the load sharing among different DERs.
In practice, the value of the virtual resistance is set to enable proportional load sharing \cite{ref2m,ref01}.

Another primary control architecture is in the form of Current Source Converter (CSC).
CSC units do not participate in output voltage regulation and are usually operated at their individual maximum efficiency point using the Maximum Power Point Tracking (MPPT) algorithm \cite{ref2m}.

\subsubsection{Secondary control}
It is well known that under decentralized primary droop control, the bus voltages vary with changes in the load demand and the power generation capacities of the renewable DERs \cite{ref2m,ref01}.
Moreover, the load is not ideally shared among different DERs due to mismatched line admittances \cite{ref6m}.
In this context, the role of the {secondary control} is to alleviate the drawbacks of the decentralized droop control and restore the bus voltages to a predefined and optimized global reference (determined by the tertiary control) and to foster proper load sharing.
This is achieved by adding correction offsets to the reference voltage control parameter of the local droop controller, see Fig.~\ref{fig:control}.
We assume that the secondary control is centralized and implemented as follows:
\begin{enumerate}
\item DER $u\in\mathcal{U}$, periodically (e.g., every $5\;\text{milliseconds}$) sends a short message to the CC with payload $[\tilde{v}_u,\;\tilde{i}_u]$ with $\tilde{v}_u$ and $\tilde{i}_u$ denoting the local measurement of the output bus voltage and current;
\item the CC computes the voltage restoration and proportional load sharing offsets $\delta x^{\text{v}}$ and $\delta x_u^{\text{c}}$ using Proportional-Integral (PI) loops;
\item the CC sends unicast packet with payload $[\delta x^{\text{v}},\delta x_u^{\text{c}}]$ to DER $u$, $u\in\mathcal{U}$;
\item upon receiving the packet, each DER uses $\delta x^{\text{v}},\delta x_u^{\text{c}}$ to correct the local droop controllers: 
\begin{equation}\nonumber
	v_u^{\star} = x_{u} + \delta x^{\text{v}} + \delta x_u^{\text{c}} - r_u i_{u},\;u\in\mathcal{U}.
\end{equation}
\end{enumerate}
After the load/generation {change}, the offsets $\delta x^{\text{v}}$ and $\delta x_u^{\text{c}}$ converge to stable values and remain fixed until the next fluctuation.
The modified droop control law provides a global voltage regulation reference and fosters proportional load sharing.
Note that only VSC DERs participate in secondary control voltage and current sharing regulation.

\subsubsection{Tertiary control}
The \emph{tertiary control} runs with significantly lower frequency than the secondary control (e.g., every $5-30$ minutes \cite{ref7m}), optimizing the performance of the system and generating the optimal references for the lower control levels.
The tertiary control objective depends on the specific application.
In our system, a generic centralized tertiary control is implemented as follows:
\begin{enumerate}
\item DER $u\in\mathcal{U}$ periodically sends a message with generic payload $z_u,u\in\mathcal{U}$ to the CC;
\item the CC solves the application-specific optimization problem and generates the optimized global/local control parameters;
\item the optimal parameters are sent to the DERs in a message with payload $q_u,u\in\mathcal{U}$;
\item the DERs follow the received CC directives and reset the local control parameters which remain valid until new ones are received.
\end{enumerate}

\begin{figure}[!tb]
\centering
{\includegraphics[width=\columnwidth]{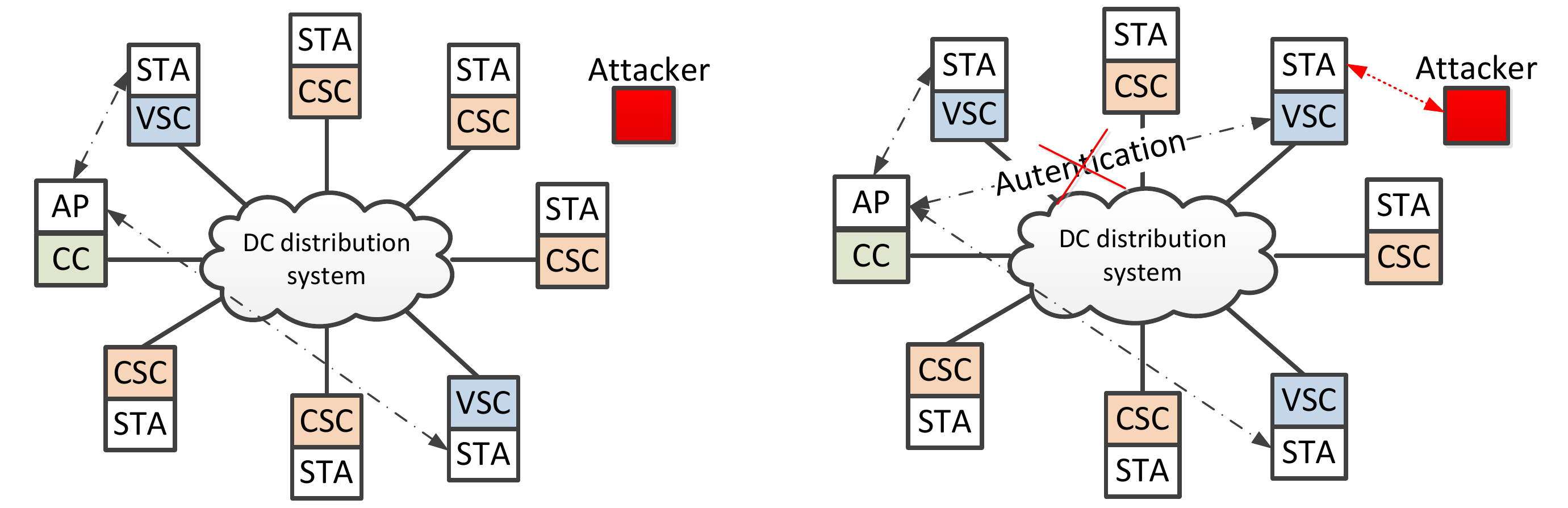}}
    \caption{Attacking the secondary control level via disabling initial handshake.}
    \label{fig:attacks}
\end{figure}

\begin{figure*}[!t]
\centering
{\includegraphics[scale=0.6]{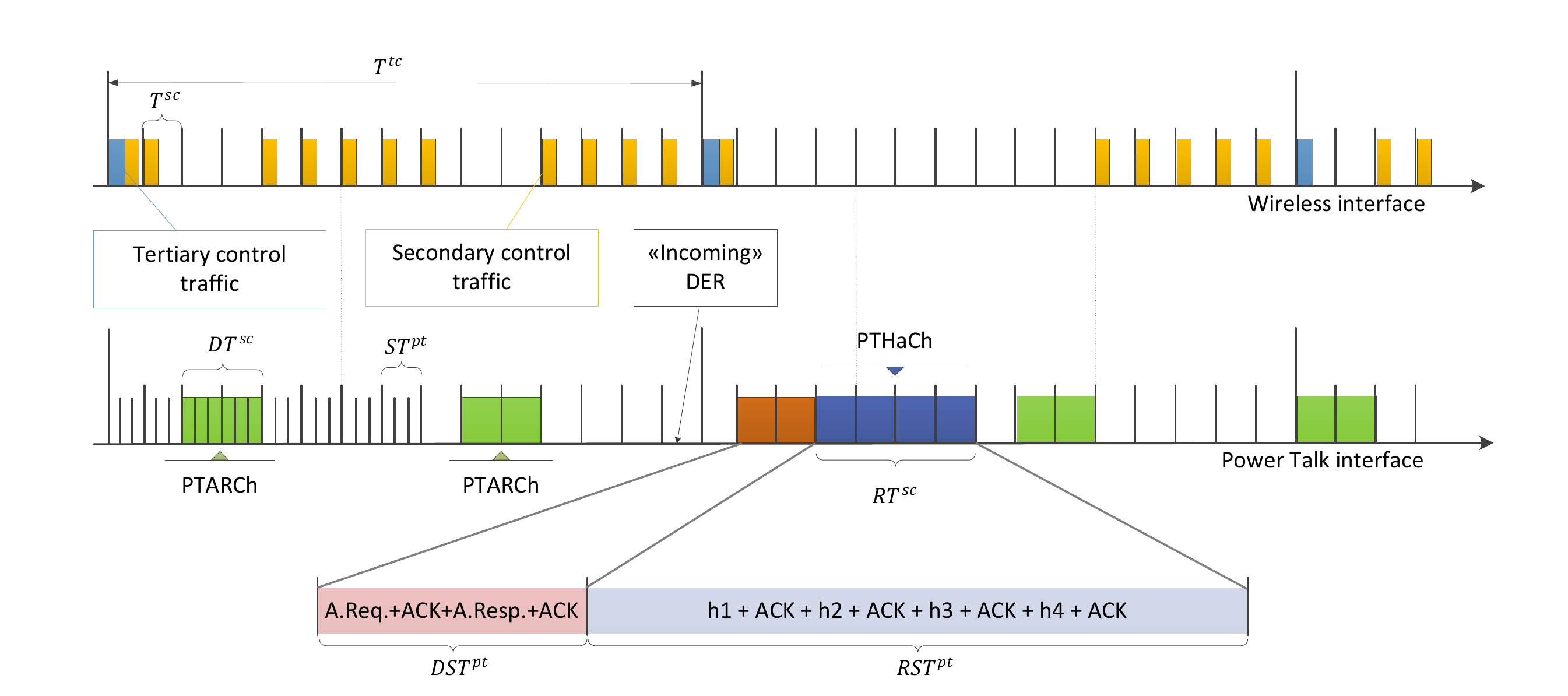}}
    \caption{Time organization of the proposed power talk-based DER authentication scheme.}
		\label{fig:axis}
\end{figure*}

\subsection{Attacks on the control system via attacks on the wireless interface}

To motivate the development of our framework, we take a closer look into two potential attacks on the secondary control.

Consider DC MG in which small subset of DERs are operated in VSC mode, while the rest are operated in CSC mode, see the right-hand side of {Fig.~\ref{fig:attacks}.}
Such primary control architectures are typical in state-of-the-art installations, where the individual renewable DERs (e.g., solar panels) are operated at their respective maximum efficiency points \cite{ref1m,ref2m}.
A known problem in such systems is the limited ability of the controlling VSC DERs to restore and stabilize the voltage around the global reference after frequent dramatic changes in the renewable power capacity of the CSC DERs; this decreases the overall system efficiency and might lead to instability \cite{ref2m}.
One possible solution is to configure the DERs with \emph{dual mode capability}, i.e., to switch between CSC and VSC modes transparently \cite{ref2m}.
Specifically, whenever the grid voltage crosses predefined voltage thresholds, portion of the CSC DERs switch to VSC primary control mode and ask the CC to join the secondary control.
To gain access to the wireless communication resources, they are initially requested to authenticate.

An attacker might attack the control system by 1) attacking the handshake and disabling the authentication, which prevents the newly switched VSC DERs from joining the secondary control, see the left-hand side of Fig.~\ref{fig:attacks}, and 2) jamming the wireless channel of one or several VSC DERs, including the CC, that participate in secondary control \cite{ref03}.
Both types of attacks result in poor voltage regulation, leading to performance degradation and potential instabilities.

The solution presented in the paper addresses the first type of attack (Fig.~\ref{fig:attacks}), by implementing the handshake over parallel powerline communication interface based on power talk, which is described in Section~\ref{sec:ptmac}. 
We also note that the power talk interface can be easily adapted to address the second type of attack, by allowing jammed units to send an alarm to the CC.
The CC can then act appropriately, e.g., by sending directives to the jammed VSCs DERs via power talk to switch back to CSC mode, and asking non-jammed CSCs to switch to VSC mode and join the secondary control.
Finally, the power talk interface can be also used as a safe and secure channel over which the VSC DERs can re-elect new CC in case the wireless interface of the current CC is under attack.
Addressing these aspects is part of ongoing work.
  
\subsection{Power Talk}
\label{sec:ptmac}

Power talk is implemented on primary droop control level, and requires the secondary control to be switched off during its operation \cite{ref3m,ref4m,ref5m}.
{VSC} DER $u\in\mathcal{U}$ modulates information into the values of the local reference voltage $x_u$ and virtual resistance $r_u$ droop control parameters, thus inducing disturbances of the output voltages.
At the same time, droop controlled DERs also observe the steady state bus voltage response.
We say that the inputs to the \emph{power talk multiple access channel} are $x_u$ and $r_u$, $u\in\mathcal{U}$, while the output observed by DER $k\neq u$ is represented through the disturbances of the output bus voltage.
The power talk channel is non-linear and requires full knowledge of the configuration of the system, which makes it particularly challenging for implementation \cite{ref3m,ref5m}.
However, the main challenge and impairment stems from the requirement for turning off the secondary control; this makes the steady state voltage susceptible to random load variations which alter the output voltage of the DERs in unpredictable manner.
Large portion of the work on power talk consist in designing viable strategies to mitigate the effect of random load changes in various communication scenario, such as one-way, broadcast, all-to-all full duplex, etc. {\cite{ref3m,ref4m,ref5m}}.
As detailed in the Section~\ref{sec:main}, here we employ a basic variant, suitable for one-way communication where only one transceiver pair is active at a time.



\section{Secure and Robust DER Authentication Protocol based on Power Talk Communication}
\label{sec:main}

Here we describe the scheme for secure and reliable power talk based authentication of a DER to the external communication infrastructure.
We label an ``incoming'' DER with $U$, and we assume that it is {connected to the MG via primary control and does not participate in upper level control}.
DER $U$ wishes to join the set $\mathcal{U}$, i.e., to actively participate in the regulation and optimization of the MG.
Prior to this, DER $U$ should be authenticated and granted permission to join the wireless networking.
The authentication is performed by the CC; 
the control architecture of the DERs engaged in the handshake is summarized in Fig.~\ref{fig:control}.

\subsection{Protocol Organization}

Fig.~\ref{fig:axis} depicts the time organization of the external wireless and power talk interfaces; {note all DERs physically connected to the system are synchronized}.\footnote{The synchronization is easy to achieve and maintain through the standard techniques used in IEEE 802.11.}
The time axis of the wireless interface is divided into secondary and tertiary control periods of durations denoted with $T^{\text{sc}}$, $T^{\text{tc}}$, corresponding to the sampling frequencies of the secondary and tertiary controllers, respectively.
Note that $T^{\text{sc}}\ll T^{\text{tc}}$.

The power talk interface is based on a \emph{periodically} repeating pool of $D\geq 1$ consecutive secondary control periods in which the secondary control is turned off, see Fig.~\ref{fig:axis}.\footnote{The secondary control is turned off to enable establishment of the power talk channel; during the off periods, the CC does not send secondary control information and the DER $u$ uses the last available $\delta x^{\text{v}}$, $\delta x_u^{\text{c}}$.}
We name the periodic pool as \emph{power talk association request channel} (PTARCh), with duration $T^{\text{sc}}D$, occurring with frequency $\frac{L}{T^{\text{tc}}}$.
The PTARCh is used by any ``incoming'' VSC DER that wants to join the upper level control, to send initial association request to the CC.
After receiving association request, the CC keeps the secondary control off and allocates additional, \emph{on-demand} pool of $R\geq 1$, consecutive secondary control periods for execution of the actual handshake, establishing the on-demand \emph{power talk handshake channel} (PTHaCh), see Fig.~\ref{fig:axis}.
Specifically, DER $U$, after physically connecting to the MG through primary control and synchronizing with the power talk interface, switches to VSC mode and waits until the next occurrence of the PTARCh.
Then, it sends association request to the CC.
The CC, after receiving the request successfully, first sends an ACK and then the response message to to DER $U$.
DER $U$ acknowledges the response which triggers the CC to allocate additional power talk resources for the rest of the handshake, namely the exchange of messages $h_1$ to $h_4$, see Fig.~\ref{fig:hs}, which happens in the PTHaCh channel of duration $T^{\text{sc}}R$, see Fig.~\ref{fig:axis}.
We assume that the duration of a single secondary control period can be expressed as $T^{\text{sc}} = S T^{\text{pt}}$, where $T^{\text{pt}}$ denotes the duration of a single power talk slot and $S\geq 1$ is an integer.\footnote{The duration $T^{\text{pt}}$ complies with the control bandwidth of the inner primary control loops, such that the system reaches steady state in a power talk slot. Typically $S\leq 10$.
}

The described protocol relies on switching off the secondary control for limited period of time, during which the system becomes susceptible to voltage deviations, required by the power talk (see subsection~\ref{sec:ptmac}).
However, this also makes the steady state voltage susceptible to load variations, representing the main communication impairment of the power talk interface, as described next.
 
\subsection{Specifics of the Power Talk PHY interface}

We employ binary power talk introduced in \cite{ref3m}.
Fig.~\ref{fig:control} gives an overview of the main functional blocks residing in the CC and DER $U$ when engaged in handshake via power talk.
Label the power talk transmitter in a power talk slot with $i$ and the receiver with $j$; when $i=U$ then $j=0$ and vice versa.
The handshake messages, represented as binary strings are mapped into binary stream of reference voltage deviations in the transmitters' droop control loop, as follows:
\begin{align}\nonumber
& 0\;\leftrightarrow\;x_{i} - \gamma,\\\nonumber
& 1\;\leftrightarrow\;x_{i} + \gamma,
\end{align}
The reference voltage deviation amplitude $\gamma$ satisfies $\frac{\gamma}{x_i}\ll 1$, see \cite{ref3m} for discussion on how to choose $\gamma$.
The deviation $\pm\gamma$ leads to deviations of the output voltage of the receiving DER:
\begin{align}\nonumber
& 0\;\leftrightarrow\;v_{j} - \Delta v_j(0),\\\nonumber
& 1\;\leftrightarrow\;v_{j} + \Delta v_j(1),
\end{align}
where $v_j$, corresponding to $\gamma = 0$.
The receiver, in each power talk slot collects $\nu (T^{\text{pts}}-\tau)$ noisy steady state output voltage samples, denoted with $\tilde{v}_j[n]$, then, it compares their average to the threshold $v_j$, and makes decision on the information bits:
\begin{align}\nonumber
 \text{if } \frac{1}{\nu T^{\text{pt}}}\sum_{n}\tilde{v}_j[n]  > v_j \text{, decide}\; 1,\\\nonumber
 \text{if } \frac{1}{\nu T^{\text{pt}}}\sum_{n}\tilde{v}_j[n]  < v_j \text{, decide}\; 0.
\end{align}
Notice that $\tau$ denotes the transient interval in which the bus reaches a steady state.

The above detection scheme is challenged by two major impairments: (i) the susceptibility of the primary control level to sporadic load variations during the periods in which the secondary control is off, and (ii) sampling noise of the converters' ADC.

The impact of load variations on power talk has been extensively studied \cite{ref3m,ref5m}.
Specifically, in the above scheme, a load change invalidates the detection threshold $v_j$, which might lead to burst of bit errors.
A simple strategy to deal with this impairment is to employ load change detection, in parallel with power talk symbol transmission/detection, in both the transmitter and the receiver \cite{ref3m}.\footnote{Both the transmitter and the receiver are able to determine whether a load change has occurred by tracking their respective output voltage levels in each power talk slot.}
After detecting load change, the transmission is paused, $M$ ``blank'' power talk slots, i.e., slots with $\gamma = 0$ are inserted and the communicating DERs to measure the new steady state output voltage level, which is used as the new detection threshold.
The viability of this technique is verified in Section~\ref{sec:results}.
Note that under this strategy, each load change increases the time necessary to complete the handshake.
We introduce the \emph{average handshake completion time} denoted with $\mu$, defined as the average time needed to complete the handshake, after the DER accesses the PTARCh.
We model the load changing process as a Poisson process with arrival rate $\lambda$ \cite{ref5m}.
Then, the following expression for $\mu$ can be derived \cite{ref5m}:
\begin{equation}\nonumber
\mu = (D+R)ST^{\text{pt}}(1-e^{-\lambda}+e^{\lambda M}).
\end{equation}
The expression shows that when the rate $\lambda$ is significantly lower than the power talk signaling rate, the dominant contributing factor to $\mu$ is the {total number of bits $(D+R)S$} of the handshake messages. 
The sampling noise of the converter follows Gaussian distribution \cite{ref8m}, i.e., $\tilde{v}_j[n]\sim\mathcal{N}(v_j + \Delta v_j(b),\sigma^2)$ \cite{ref5m}.
The \emph{bit error rate (BER)} can be calculated as:
\begin{equation}\nonumber
P_e = 1 - \frac{1}{2}\text{erf}\bigg(\frac{\Delta v_j(1)}{\sigma\sqrt{2}}\bigg) - \frac{1}{2}\text{erf}\bigg(\frac{\Delta v_j(0)}{\sigma\sqrt{2}}\bigg).
\end{equation}
An appropriate error correction code can be employed to protect the handshake messages against noise-related errors.

\section{Results}
\label{sec:results}

We simulate a simple DC MG using PLECS\textsuperscript{\textregistered{}}.
There are $U=6$ DER units, connected to a single bus in parallel via distribution lines with resistances $0.2\,\Omega$.
The aggregate resistive load of $1.5\,\Omega$ is also connected in parallel to the bus.
We use $T^{\text{sc}} = 5\,\text{ms}$ and {$T^{\text{tc}} = 5\,\text{min}$}.
The reference voltages of the VSC DERs and the global grid voltage reference are equal and set to $x_u = v^{\star} = 48\,\text{V}$.
The virtual resistances of all VSC DERs are set to $0.2\,\Omega$ and are kept fixed.
{Given the above system parameters, the transient response time of a step change of the reference voltage is estimated to be $\tau = 2.35\,\text{ms}$}.
The wireless interface is IEEE 802.11n.
Using Wireshark and a compatible Wi-Fi interface, we capture the handshake messages and map their binary versions directly onto the power talk interface.
The length of the messages in bytes is 114 for the Association Request, 93 for the Association Response, 177 for \emph{h1}, 177 for \emph{h2}, 211 for \emph{h3}, 155 for \emph{h4} and 32 for ACKs.
The ADC sampling noise standard deviation is denoted with $\eta = 8.58 \cdot 10^{-2}$.
Thus, the standard deviation of the power talk samples in each power talk slot can be calculated as $\sigma=\frac{\eta}{\sqrt{\nu (T^{\text{pt}}-\tau)}}$.

\begin{figure}[!tb]
\centering
\subfloat[Output currents of the $6$ DGs.]{\includegraphics[width=\columnwidth]{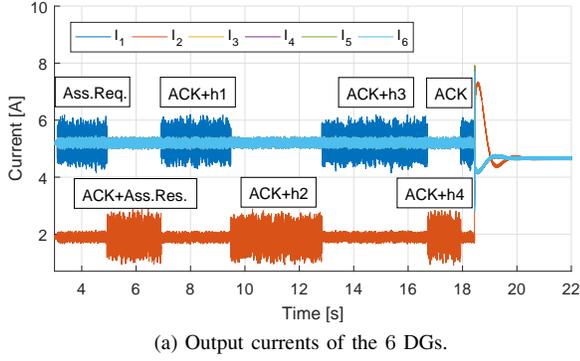}\label{fig:procedurea}}
\hfil
\subfloat[Grid voltage.]{\includegraphics[width=\columnwidth]{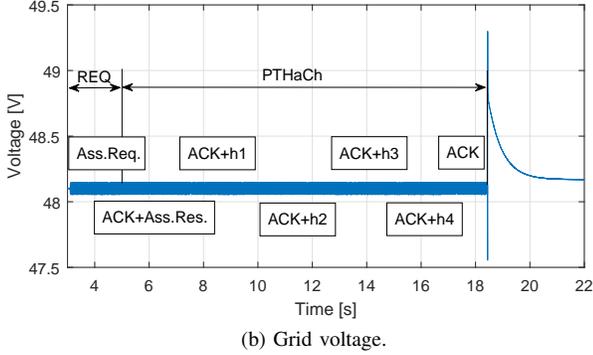}\label{fig:procedureb}}
    \caption{Realization of an authentication process.}
    \label{fig:procedure}
\end{figure}

\begin{figure}[!tb]
\centering
\subfloat[Output currents of the $6$ DGs.]{\includegraphics[width=\columnwidth]{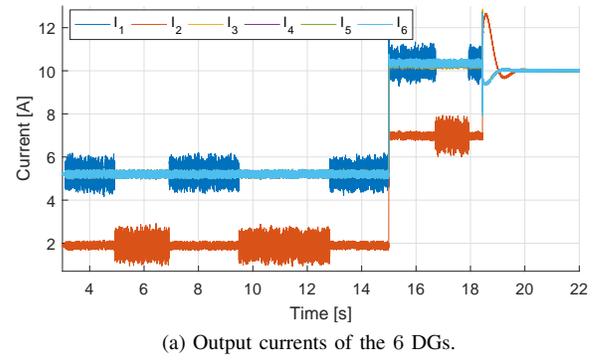}}
\hfil
\subfloat[Grid voltage.]{\includegraphics[width=\columnwidth]{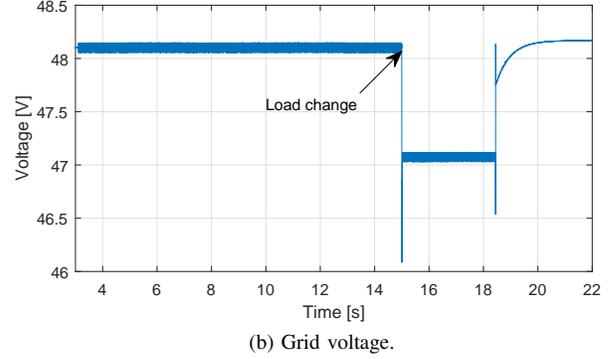}}
    \caption{Realization with a load variation.}
    \label{fig:loadchange}
\end{figure}


We first demonstrate and verify the technical feasibility of the proposed authentication scheme.
We use uncoded power talk modulation, i.e., we directly map the handshake binary representations of the messages to the power talk interface.
The CC receives the association request packet from DER $U$ in the PTARCh starting at $t=3$ s.
Fig.~\ref{fig:procedure} depicts the output currents of the DERs and grid voltage during and after the handshake.
Before authorization, it is clear that DER $U$, being connected to the MG only through primary control, does not participate into secondary control regulation and its output current is not contributing proportionally to the load.
After the handshake ends and the secondary control becomes switched on, we observe that the output current of DER $U$ quickly becomes aligned with the output currents of the other DERs, verifying that the authentication was successful and that DER $U$ joined the secondary control.

Fig.~\ref{fig:loadchange} illustrates a handshake during which a load change occurs, in $t=15\,\text{s}$.
After resetting the detection threshold, the handshake resumes.
Upon successful authorization to DER $U$ and turning the secondary control on, we observe that the load is proportionally shared and the voltage is restored to its global reference.
Taking into account that the load changes rather infrequently compared to the signaling rate of power talk {\cite{ref3m,ref5m}}, sporadic load changes in the power talk channels can be efficiently mitigated using the detection reset strategy with negligible impact on the overall duration of the power talk periods.
Moreover, the impact of any load change that occurs during the void PTARChs, i.e., when no DER sends association request, will be quickly eliminated after the secondary control is switched on, restoring the voltage to its global reference and fostering proportional current sharing.

\begin{figure}[!tb]
\centering
{\includegraphics[width=\columnwidth]{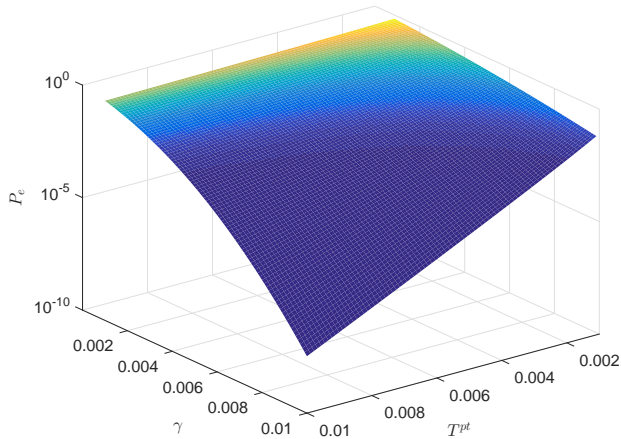}}
    \caption{Bit error rate of the power talk interface as function of $\gamma$ and $T^{\text{pt}}$.}
    \label{fig:ber}
\end{figure}

Figs.~\ref{fig:procedure} and \ref{fig:loadchange} also illustrate that the handshake can take up several seconds to complete via power talk which is a result of the fundamental fact that power talk is a narrowband communication technique.
During the handshake period, if a load change occurs, the operating point will be suboptimal as the system is governed only by the primary control.
However, we note that, as a proof of concept, the figures are derived using the actual handshake messages from IEEE 802.11 standard, which is a high-rate wireless interface with the symbol durations that are only fractions of microsecond.
If the structure and length of the handshake messages and possibly the complete handshake procedure were redesigned to match the features of the power talk interface, the duration of the procedure could be significantly shortened.
This would also significantly reduce the fraction of time during which the system is susceptible to load variations.

Finally, Fig.~\ref{fig:ber} depicts the BER as a function of the reference voltage perturbation amplitudes $\gamma$ and the duration of the power talk slot $T^{\text{pt}}$.
Clearly, the impact of the sampling noise becomes negligible as $\gamma$ and $T^{\text{pt}}$ increase.
Remarkably, already with $\gamma = 0.01$ (i.e., $0.02\%$ of $v^{\star}$) volts and $T^{\text{pt}}=0.01\,\text{s}$, the BER falls below $10^{-7}$, implying that the power talk interface can be used in conjunction with simple error correction codes with negligible redundancy and low complexity.

\section{Conclusion}
\label{sec:conc}

If the MG control is simply placed on top of an existing communication technology, it will also inevitably inherit the corresponding security threats.
According to this paradigm, adopted by previous works related to MG security, the system should include countermeasures against any known security vulnerability, to enable reliable and secure operation of the control system.

In this paper we show that if the communication is secured using a non-conventional channel, i.e., the grid itself, it becomes safe against 
traditional cyber-attacks.
In particular, we modeled and demonstrated an authentication scheme for IEEE 802.11 systems based on power talk.
Its robustness stems from the fact that the initial handshake can be observed and altered by only being physically connected to the grid. 
Moreover, power talk can play a pivotal role in the overall security, e.g., by reporting communication outages of the primary channel and by distributing its encryption keys.
These topics are included in our ongoing work.


\section*{Acknowledgment}

{The work presented in this paper was supported in part by EU, under grant agreement no. 607774 ``ADVANTAGE''.}



%
\nocite{*}
\bibliographystyle{IEEEtran}
\bibliography{IEEEabrv,refs}

\end{document}